\documentclass[referee]{raa}           

\usepackage{graphicx,times}
\usepackage{natbib}
\usepackage{physics}
\usepackage{amssymb,amsmath}
\usepackage{indentfirst}
\usepackage{xcolor}
\bibpunct{(}{)}{;}{a}{}{,}
\usepackage[pagebackref=true]{hyperref}

\newcommand{\mr}[1]{\mathrm{#1}} 

\defcitealias{Ruderman+etal+1975}{RS75} 

\begin{document}

   \title{Pulsar Sparking: What if mountains are on the surface?}

 \volnopage{ {\bf 2024} Vol.\ {\bf X} No. {\bf XX}, 000--000}
   \setcounter{page}{1}

   \author{Zi-Hao Xu \inst{1}, Wei-Yang Wang \inst{2}, Shun-Shun Cao \inst{1}, Ren-Xin Xu \inst{1}
   }

   \institute{ Department of Astronomy, Peking University, Beijing 100871, 
China;\\
\and 
School of Astronomy and Space Science, University of Chinese Academy of Sciences, Beijing 100049, China
{\it wywang@ucas.ac.cn}\\
\vs \no
   {\small Received 20XX Month Day; accepted 20XX Month Day}
}

\abstract{A numerical framework to calculate the height and potential of the vacuum inner gap of pulsars is presented here.
The results demonstrate that small mountains on a pulsar's polar cap tend to significantly influence the properties of the inner vacuum gap, making it easier for sparks to form.
In this scenario, the magnetospheric activity observed from the pulsars PSR J0250$+$5854 and PSR J2144$-$3933 which lie below the traditional pulsar death line, and some single-pulse modulation phenomena could also then be understood.
Furthermore, the presence of small mountains should depend on the puzzling state of supranuclear matter inside pulsars.
In order to sustain stable mountains on the surface, pulsars might be made of solid strangeon matter, which is favoured by both the charge neutrality and the flavour symmetry of quarks.
}
\keywords{pulsars: general  --- radiation mechanism: nonthermal --- pulsars: individual (J0250+5854, J2144-3933) --- stars: dense matter state 
}

   \authorrunning{Z.-H. Xu et al. }            
   \titlerunning{Pulsar surface mountain}  
   \maketitle

%
\section{Introduction}           
\label{sect:intro}

The state of the supranuclear dense matter inside a pulsar has long been a controversial topic.
Because of the complexity of non-perturbative quantum chromodynamics (QCD) ~\citep{Dosch+1994,Fischer+2006,Degrand+2006}, it has been almost impossible to predict theoretically the inner structure of pulsars until now.
Traditional wisdom suggests that pulsars \citep{Hewish+1968,Gold+1968} are neutron stars, superficially anticipated by \cite{Landau:1932uwv} but hypothesized by \cite{Baade+1934} and \cite{Oppenheimer+1939}.
However, following the establishment of the standard model of particle physics, it has been conjectured that pulsars are composed of strange quark matter~\citep{Witten+1984,Alcock+1986}, or even strangeon\footnote{A strangeon is actually a strange quark cluster~\citep{Xu+2003} containing an equal number of three light-flavor quarks \citep{Xu+2019,Xu+2021}. It is a nucleon-like bound state, but with strangeness.} matter~\citep{Lai+2017,Zhang+2023}. 
A general review of strange matter has recently been presented~\citep{2025arXiv251101146X}. Furtheremore, a pulsar's radiative properties would depend on its state of matter~\citep{Meszaros+1992,Adelsberg+2006,2022ASSL..465..281B}.
For example, a pulsar's ability to emit radio emissions could be relevant to the binding energy of the particles on its surface and its geometry\, and a bare surface on a strange star could help to solve the ``binding energy problem'' in the Ruderman-Sutherland model~\citep{Xu+1999}.

In fact, this classical vacuum inner gap model can principally explain the diversity of pulsars’ radio emissions, where the vacuum inner gap discharge (called ``sparking'') occurs in the polar cap region (\citealt{Ruderman+etal+1975}, hereafter \citetalias{Ruderman+etal+1975}).
In this model, a strong electric field parallel to the magnetic field in the inner gap accelerates positrons to have relativistic kinetic energies, enabling curvature radiation that produces high-energy photons \citep{Sturrock+1971}.
These gamma-ray photons subsequently undergo magnetic pair production \citep{Schwinger+1951,Adler+1971} in the pulsar's intense magnetic field, initiating a pair cascade \citep{Daugherty+1983}. 
The resulting avalanche of positrons creates a discharge responsible for coherent radio emissions \citep{Sturrock+1971,Tademaru+1971}.
Crucially, when the maximum potential from unipolar induction becomes insufficient to sustain this pair cascade, the pulsar transitions to a radio-quiet state.

The \citetalias{Ruderman+etal+1975} model is frequently challenged by observations of two ``dead" pulsars, PSR J0250+5854 (\citealt{Tan+etal+2018}) and PSR J2144-3933 (\citealt{Young+etal+1999}).
These pulsars have radio emissions even if they lie below the predicted ``death line".
However, as demonstrated in this article, if the surface is rugged, e.g., with small ``mountains'' or ``zits'' \citep{Xu+2023}, the parallel electric field near the surface might be enhanced, enabling more efficient positron acceleration.

Several theoretical models have already been proposed to re-initiate the cascade pair production explain those unexpected radio pulses from the stars below the pulsar death line. \cite{Chen+1993} argue that a more complicated magnetic field structure can make the pair production more likely to take place due to stronger curvature radiation. With general relativity frame dragging taken into account, inverse Compton scattering (ICS) induced space-charge limited flow (SCLF) model can sustain strong pair production in some long period pulsars \citep{Zhang+2000}. However, the pulsars of interest are located beyond the death valley predicted by the ICS-induced vacuum gap model. Although the ICS-induced SCLF model successfully explains long-period pulsars, it fails to account for the subsequently discovered millisecond pulsars without additional assumptions \citep{Zhang+2000}. Pulsars with small inclination angles are also supposed to be more capable to sustain the emission with long period, compared with usual pulsars \citep{Spitkovsky+2006}. A population synthesis based model by \cite{Szary+2014} shows the plasma generating radio emission is produced under similar condition regardless of different pulsars, enabling pulsars below the death line to keep shining.

Nevertheless, in this paper we develop a numerical method to calculate how polar cap surface ruggedness or mountains influence the vacuum inner gap within the \citetalias{Ruderman+etal+1975} model.
This method quantifies the ability of such mountains to modify the potential drop, which fundamentally determines the threshold for spark discharge processes. Meanwhile, the presence of small mountains may also provide possible explanation for some other peculiar radiation properties, such as subpulse, unbalanced sparking, mode switching and even swooshing (See Section~\ref{sect:discussion} for details).

The structure of this article is as follows. In Section~\ref{sect:method}, we propose a procedure to systematically estimate the height of the inner gap layer and the potential drop across the gap region required to trigger the cascade process.
In Section~\ref{sect:results}, we apply the procedure to calculate the influence of an ideal mountain on the potential drop across the inner gap region for the two ``dead" pulsars, PSR J0250+5854 and PSR J2144-3933 as mentioned above. In Section~\ref{sect:discussion}, we discuss the physical implications of these results and analyze limitations and speculations about how to test our model.

\section{calculating the inner gap height and potential drop} \label{sect:method}

\subsection{
Accelerating field of the inner gap} 
\label{sect:accfield}

The corotation of magnetosphere and the star induces the Goldreich Julian charge distribution in the static magnetosphere \citep{Sturrock+1971}. However, the outflow of charges near the pulsar's light cylinder removes positive ions from the magnetosphere (\citealt{Cheng+1976}), while the  binding energy of surface ions ($\sim14$ keV; \citepalias{Ruderman+etal+1975}, or lower \citep{Hillebrandt+1976,Flowers+1977,Kossl+1988,Lai+2001}) prevents effective charge replenishment. So for a certain type of pulsar model, if the binding energy of surface ions is high enough, there will be charge starvation above the polar cap.
This charge starvation inevitably leads to the formation of a vacuum inner gap,  where the electric field holds a component parallel to the magnetic field ($\vb*{E}\vdot \vb*{B}\neq 0 $), and manages to accelerate electrons and positrons .

The potential drop across the gap can be approximated by $\Delta V \simeq \Omega B h^2/2$  \citepalias{Ruderman+etal+1975} with small gap height. If the surface of the polar cap region is absolutely flat, i.e. for the 1D approximation
one can imagine the gap as a parallel capacitance with zero potential on the lower surface and zero field on the upper surface, so the parallel electric field is
\begin{equation}
    E_\parallel (z)=2\frac{\Omega B}{c}(h-z), \label{eq:zero}
\end{equation}
which depends on the height of the inner gap.

\subsection{
The ICS process and mean free paths of positrons and photons
}
\label{sect:ics_mfp}

The characteristic height of the gap can be approximated by the mean free path of relativistic positrons (electrons) and the emitted high-energy photons within the strong magnetic field.
While \citetalias{Ruderman+etal+1975} attributes the dominant radiation mechanism to curvature radiation (CR), in which accelerated positrons convert their kinetic energy into high-energy photons, we argue that inverse Compton scattering (ICS),  first  applied to pulsar radiation mechanism by \cite{Sutherland+1979, Xia+etal+1985} and \cite{Daugherty+1989}, can dominate the production of high-energy photons due to the higher efficiency compared to the CR process\citep{Arons+1983,Zhang+etal+1996,Zhang+etal+1997,Xu+2000}.

For a typical magnetic field \(B = 10^{12}\:\mr{G}\), the energy of ICS photons can be \(\hbar \omega_s \simeq 4\:\mr{GeV}\) with \(\gamma \sim 10^5\), compared to the \(300\:\mr{MeV}\) CR photons requiring \(\gamma \sim 10^6\). 
This efficiency stems from the ICS photon energy scaling as \(\hbar \omega_s = 2\gamma \hbar e B/(m_e c)\), which depends directly on the magnetic field strength rather than the electron¡¯s trajectory curvature. Thus the ICS cross section is greatly enhanced in strong magnetic fields due to resonant scattering. As a result, the potential drop required by sparks is brought down by the higher radiation efficiency, releasing the binding energy difficulty from 10 keV to several keV\citep{Zhang+etal+1996}. Therefore, ICS process is more likely to be the dominant way to produce high energy photons and ultimately determine the characteristic height of the inner gap.

\begin{figure}[ht]
    \centering
    \includegraphics[width=0.7\linewidth]{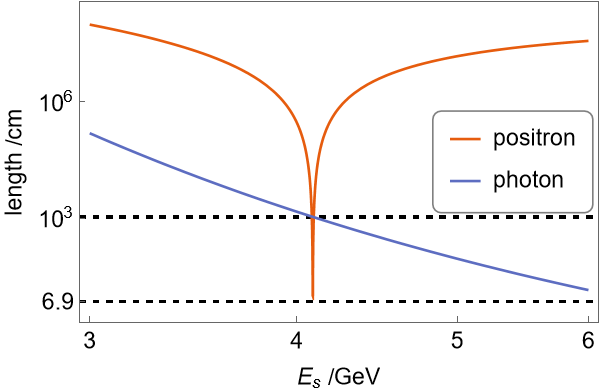}
    \caption{The relation between the mean free path of positrons and photons in a strong magnetic field environment of the gap to the upscattered photons energy respectively. The sharp dip of positron mean free path corresponds to the resonance condition $\epsilon_s=2\gamma\epsilon_B$. In the figure, $B=10^{12}\:\mr{G},P=1\:\mr{s}, T=10^6\:\mr{K},\gamma=10^5$ are adopted. }
    \label{fig:free-path}
\end{figure}

In the case of the ICS process, the inner gap can be divided into two regions. Positrons are accelerated in the lower region, while photons propagate in the upper region. Therefore, we define the discharge condition as $h = l_e + l_p $, where $l_e(\gamma)$ and $l_p(\gamma)$ are the mean free paths of the positrons and photons, respectively. The calculation of the mean free paths of high-energy photons mainly follows the methods introduced by \cite{Erber+etal+1966}, except for a slight difference in the formula used for the purpose of better approximation (See Appendix~\ref{apx:mfp_ph} for details). For the mean free path of electrons and positrons, we follow the method used in \cite{Dermer+etal+1990} and focus on the point of resonance ( See details in Appendix~\ref{apx:mfp_el}). 
Combining the calculation of photon and positron mean free paths, we can get value of both mean free paths at different energy of the scattered photons. The results is shown in Figure~\ref{fig:free-path}.

\subsection{The parallel electric field with a small mountain}

Figure~\ref{fig:free-path} shows that under typical pulsar conditions, the photon means free path (\(\sim 10^3\:\mr{cm}\)) vastly exceeds that of positrons (\(1\text{--}10\:\mr{cm}\)) at the resonant point ($\epsilon_s = 2\gamma \hbar \epsilon_B$). This disparity allows us to approximate the parallel electric field acting on a newly released positron (before its first ICS collision) as constant: \(E_\parallel(0) = 2\Omega Bh/c\). Although the ICS process is more efficient in producing high energy photons and terminating the gap, the death line is model dependent. In the framework of the vacuum gap model, there are still pulsars below the death line based on ICS process \citep{Zhang+2000}. Thus we try to explore whether the stellar surface mountains may play an important role in reviving those stars and explain other peculiar observed facts.

The height of polar cap mountains is likely constrained to \(\lesssim 1\:\mr{cm}\), as higher mountains would emit gravitational radiation strong enough to drain the rotational energy of the pulsar, but more explicit upper limit for the height of mountains on the pulsars is unavailable so far due to insufficient detector sensitivities \citep{Gittins+2024, Sieniawska+2021}.
Consequently, the mountains' influence on the electric field is localized near their vicinity, while the field distribution at larger distances asymptotically converges to Equation~\eqref{eq:zero}.

The electric field diverges at mountain peaks due to extreme curvature effects, dramatically enhancing positron acceleration in these regions. This field enhancement reduces both the positron and photon mean free paths, consequently lowering the threshold potential required to initiate pair cascade processes. To quantitatively analyze this phenomenon, we first isolate the mountain-dominated region from the complete gap geometry.
Given the rotation period, surface magnetic field, and surface temperature, the gap height can be expressed as a function of the Lorentz factor gained by positrons right before their ICS collision with thermal photons:
\begin{equation}
     h(\gamma,\epsilon_B)=l_e (2\gamma \epsilon_B, \epsilon_B,\gamma)+l_p (2\gamma\epsilon_B,\epsilon_B) , \label{eq:height}
\end{equation}
where Lorentz factor is determined by the equation below
\begin{equation}
    \gamma m_e c^2=e E_\parallel (0) l_e (\gamma). \label{eq:lorentz}
\end{equation}
The parallel electric field is taken as $E_\parallel (0)$, which is actually felt by the positrons near the star surface.

If there are no small mountains, the surface electric field is $E_\parallel (0)=2 \Omega B h(\gamma)/c$.
Combined with Equation~\eqref{eq:height} and Equation~\eqref{eq:lorentz}, we arrive at the self-consistent solution to the height of the gap with $\gamma$ determined using the quasi-Newton method.
The potential drop across the gap is given by
\begin{equation}
    \Delta V_0 =\frac{\Omega B}{c} h^2 .
\end{equation}

For typical pulsar parameters ($B=10^{12}\:\mr{G}, P=1\:\mr{s}, T=10^6\:\mr{K}$), we get $\gamma_0=4.9\times 10^4, h_0=h(\gamma_0)=3299.17\:\mr{cm} $, which is consistent with the results in \cite{Zhang+etal+1997}. The relatively small Lorentz factor comes from the assumption that only the initial accelerating path of positrons plays the decisive role in constraining the inner gap height.

\begin{figure}
    \centering
    \includegraphics[width=0.8\linewidth]{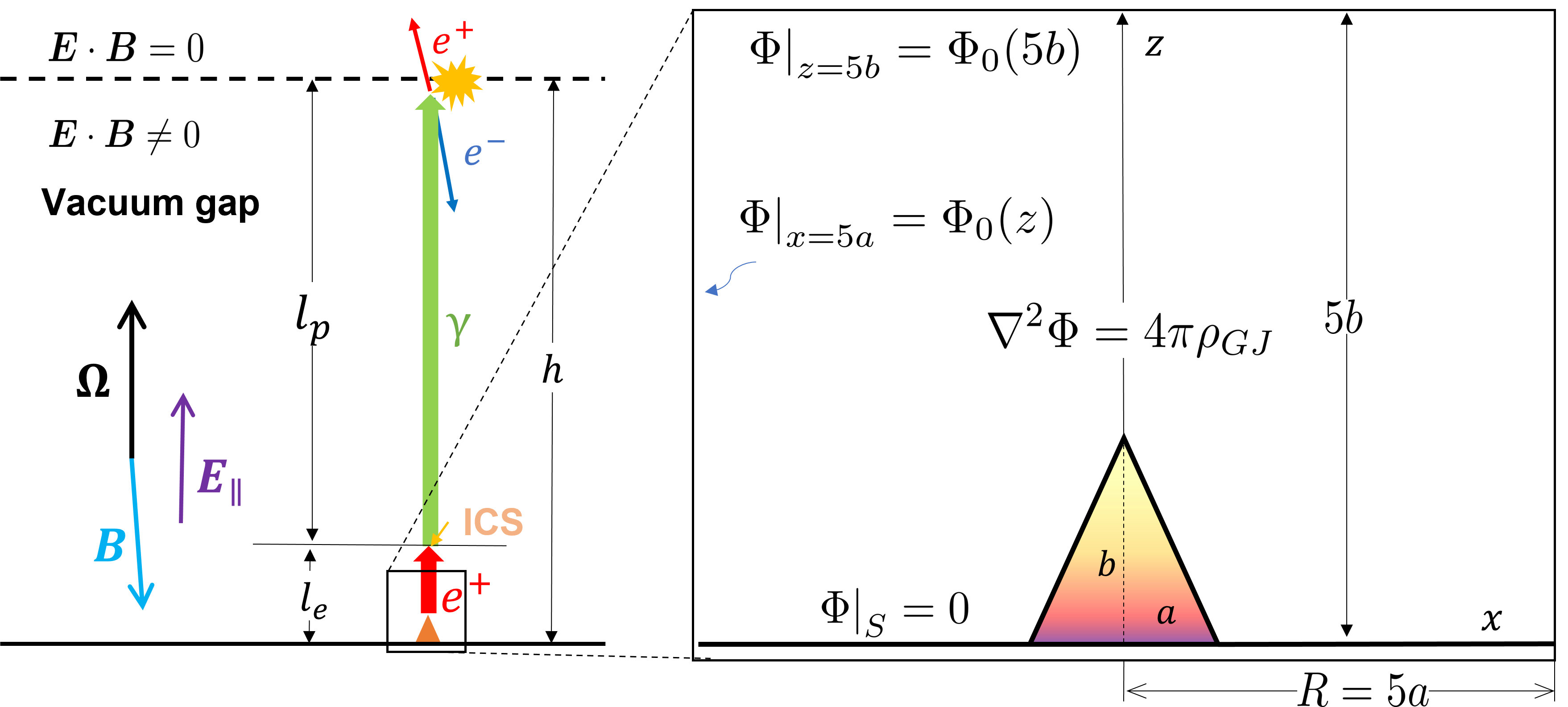}
    \caption{The electromagnetic field environment and mean free path of positrons and photons. The enlarged inset show the cylindrical vicinity of mountains with equation in the bulk and boundary condition assigned on each surface.}
    \label{fig:setup}
\end{figure}

If there is a small mountain in the polar gap region, then the actual surface electric field felt by the positrons is $E_\parallel (0) > 2 \Omega B h (\gamma)/c$. We can obtain the value by solving the effective divergence equation in the corotating frame of the pulsar with a quasi-static magnetic field in it

\begin{equation}
    \nabla \cdot \vb*{E}=4\pi (\rho-\rho_\mr{GJ}) .
    \label{eq:basic}
\end{equation}
Consider the mountain as a conic with radius $a$ and height $b$, giving a steepness of $\eta = b/a$. We choose a stellar surface as the zero potential reference and $\vb*{E}=-\nabla \Phi $. We also have $\rho=0 $ in a cylinder with radius $R=5a$ and height $H=5b$ since the net charge is zero in the gap. The boundary condition is set as

\begin{equation}
    \Phi(x,z)=\Phi_0(z)=2\pi \rho_{\rm GJ}z(z-2h_0), \quad \text{when} \quad x=R\; \text{or}\; z=h_0 , 
\end{equation}
representing the asymptotical behavior of the solution. The model sketch map is shown in Figure~\ref{fig:setup}

The second-order partial differential equation is

\begin{equation}
    \frac{1}{x} \pdv{x} \qty(x \pdv{\Phi}{x})  + \pdv[2]{\Phi}{z} = 4\pi \rho_\mr{GJ} . \label{eq:pdeeq}
\end{equation}

We solve Equation~\eqref{eq:pdeeq} by using the finite element method (FEM) on a particularly dense mesh near the mountain and obtain the result $\Phi(x,z)$.
The obtained potential distribution $\Phi(x,z)$ is then differentiated along the $z$-direction to determine the parallel electric field component $E_\parallel(x,z) = -\partial\Phi/\partial z$.

\begin{figure}
    \centering
    \includegraphics[width=0.8\linewidth]{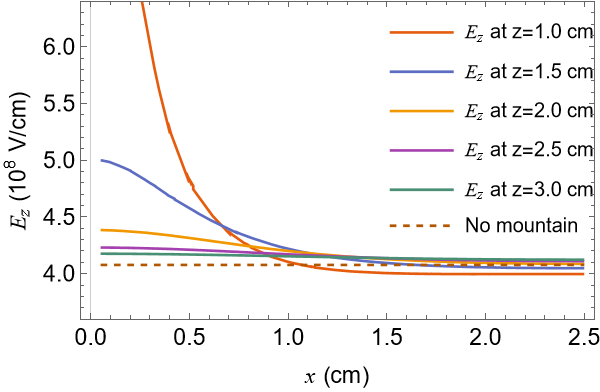}
    \caption{The distribution of the parallel electric field $E_z$ with respect to the distance to the center of the mountain, i.e. the $x$ coordinate, at different height above the stellar surface. The solid lines are for the scenario with a mountain of height $b=1\:\mr{cm}$ and $\eta = 2$, while the dashed line represents the scenario with no mountains. The influence of the mountain on the parallel electric field distribution is localized in the region with distance smaller than half the mountain radius from the origin.}  
    \label{fig:Ez-x}
\end{figure}

The numerical results show that the parallel electric field within the vicinity of the mountain top is significantly enhanced by the mountain. Also, in Figure~\ref{fig:Ez-x}, we see that the deviation from the non-mount scenario Equation~\eqref{eq:zero} is only obvious within the small region $a/2$ away from the mountain peak.
 
Hence, we substitute $E_\parallel (0)$ in Equation~\eqref{eq:lorentz} with the mean value of $E_\parallel$ within this small region:
\begin{equation}
    \bar{E}_\parallel = \frac{8}{\pi a^3}\int_0^{a/2} \dd{z} \int_0^{a/2}\dd x \cdot x E_\parallel (x,z) .
\end{equation}
The integral is calculated numerically using the first Chebyshev polynomials, and then we repeat the process of solving  Equation~\eqref{eq:height} and Equation~\eqref{eq:lorentz} simultaneously. After that, the Lorentz factor $\gamma_\text{m}$ and the height $h_\text{m}$ of the gap region with a small mountain are determined. Finally, the potential drop of the gap region is
\begin{equation}
    \Delta V_\text{m}=\frac{\Omega B}{c} h_\text{m}^2 , \label{eq:mount-potential}
\end{equation}
because electrical field is almost Equation~\eqref{eq:zero} in most part of the gap region. The subscription ``m'' in Equation~\eqref{eq:mount-potential} indicates the scenario with a small mountain in the polar cap region. 

\begin{figure}
    \centering
    \includegraphics[width=0.7\linewidth]{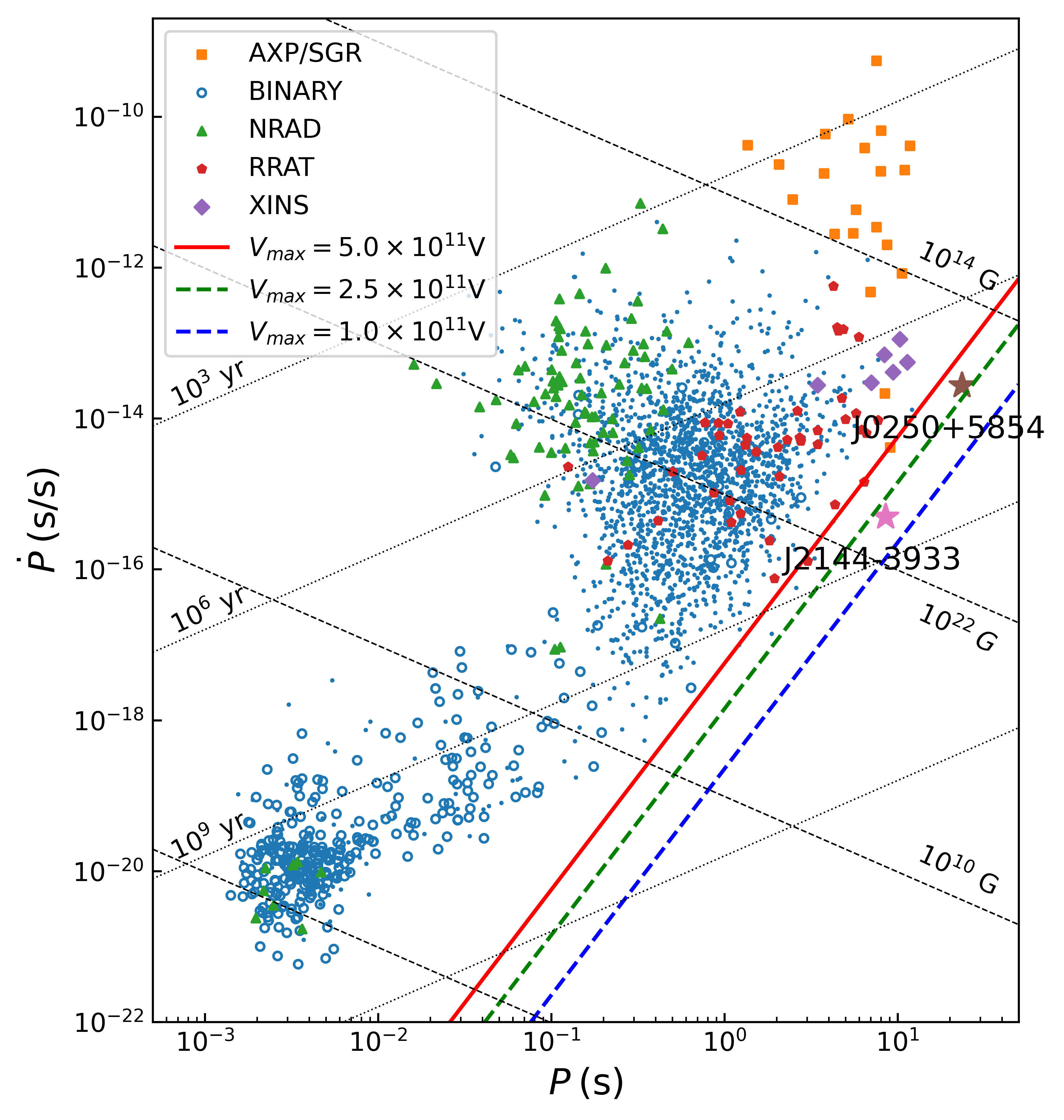}
    \caption{The $P$-$\dot{P}$ diagram of observed pulsars \citep{Manchester+etal+2005} with different categories of pulsars marked by points of different shapes. The red line is the "death line" without mountains, characterized by the maximum unipolar potential difference $5\times 10^{11}\:\mr{V}$. The green dashed line represents the death line with unipolar potential $V_\text{max} = 2.5\times 10^{11} \:\mr{V}$, i.e. mountain steepness $\eta =2$; the blue dashed line is the death line with $V_\text{max} = 1\times 10^{11}\:\mr{V} $, i.e. $\eta =6$}
    \label{fig:ppdotm}
\end{figure}

For pulsars below the death line demonstrated in Figure~\ref{fig:ppdotm}, $\Delta V_0<V_\text{max}$, where $V_\text{max}$ is the maximum potential drop that unipolar induction can produce, 
\begin{equation}
    V_\text{max}=\frac{\Omega B}{2c} r_p^2 \; ,\qquad r_p=R_s\sqrt{\Omega R_s/c}  ,\label{eq:Vmax}
\end{equation}
where $r_p$ is the radius of polar cap region.
If a mountain is present in the polar region, the spark's required potential drop decreases. When the mountain's steepness ratio $\eta = b/a$ exceeds a threshold value ($\Delta V \leqslant V_\text{max}$), spark discharges can occur.
Therefore, we can account for the radio emission by ``dead" pulsars still within the framework of \citetalias{Ruderman+etal+1975}, with the ICS process taking place of curvature radiation as a producer of high-energy photons. 

Consider a mountain with height $b=1\:\mr{cm}$ and steepness parameter $\eta = 2$. Such a topographic feature reduces the potential drop across the gap layer by a factor of 2. In contrast, a more  steep mountain with $\eta= 6$ diminishes the potential drop near the mountain to merely 20\% of its original value.

 Pulsar death line is inferred from the points $(P, \dot{P})$ on the P-Pdot diagram which satisfy the condition that maximum unipolar potential drop equals the potential drop across the vacuum inner gap. Given a parametric region on $P$-$\dot{P}$ diagram below the original pulsar death line, the stars falls into this region cannot sustain the pair production to form sparks. But the lowered potential drop will make this parametric region gain the ability to trigger sparks, i.e. this parametric region moves to the upper side of the death line, or more precisely (the $P$ and $\dot{P}$ parameters remain unchanged), the death line is shifted downwards. 
Therefore, the condition for sparks is relaxed for $V_\text{max} = \Delta V$ and the pulsar death line is shifted to the lower right corner of the $P$-$\dot{P}$ diagram due to the presence of the mountain in the polar cap region, as is shown in Figure~\ref{fig:ppdotm}.

To show more exactly how the death line changes, the potential drop $\Delta V$ and maximum unipolar potential ($V_\text{max} $)  can be regarded as the function of ($P, \dot{P}$), and find their intersection line on $P $ - $\dot{P} $ diagram, that is the death line by definition, but it is computationally intensive with mountains taking into consideration. However, since the concept of death line is merely a rough indicator of whether the a pulsar can have radio pulses or not, and the critical value chosen for $ V_\text{max} $ is something put by hand, the exact calculation may not bring us closer to the reality. 

\section{The revival of two pulsars}
\label{sect:results}

Differences in mountain height and steepness, as well as the location of the mountain in the polar cap region, can all contribute to a different potential drop across the gap.
It is necessary to take into account the complex configuration of the electromagnetic field within the inner gap to determine the influence of the mountain location measured by the polar angle $\Theta$.

\begin{figure}
    \centering
    \includegraphics[width=0.7\linewidth]{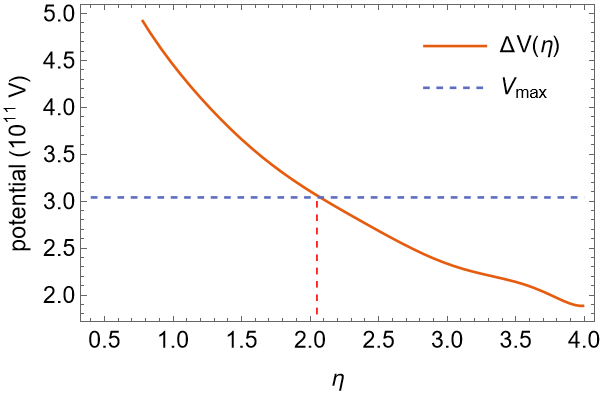}
    \caption{Orange solid line show the decreasing of the potential drop required by PSR J0250+5854 to form a spark as the  mountain steepness increasing, with blue horizontal dashed line marking the maximum potential produced by unipolar induction. When $\eta>2.1$, it hold that $\Delta V<V_\text{max}$}
    \label{fig:star1}
\end{figure}  

PSR J0250+5854 was discovered by LOFAR (LOTAAS) in 2017 (\citealt{Tan+etal+2018}), with a rotation period of $23.5\:\mr{s}$.
The surface magnetic field is referred to $2.56\times 10^{13}\:\mr{G}$ and the rotation energy loss rate is $8.2\times 10^{28}\:\mr{erg/s}$.
Considering a uniformly magnetized sphere and using Equation~\eqref{eq:Vmax}, then the maximum potential difference it can provide through magnetic unipolar induction is $ V_\text{max}= 3.04 \times 10^{11}\:\mr{V}$. If the stellar surface is assumed to be strictly flat, the potential drop required to cause a sparking discharge can be estimated as
$\Delta V_0 = 8.99 \times 10^{11} \:\mr{V}$, which exceeds the maximum possible potential difference and therefore a spark cannot be formed. However, $\Delta V$ decreases in the presence of a hill in the polar region, as shown in Figure~\ref{fig:star1}. When $\eta$ exceeds
2.1 making $\Delta V$ smaller than $ V_\text{max} $, sparks can happen, which explains the presence of radio emission.

\begin{figure}
    \centering
    \includegraphics[width=0.7\linewidth]{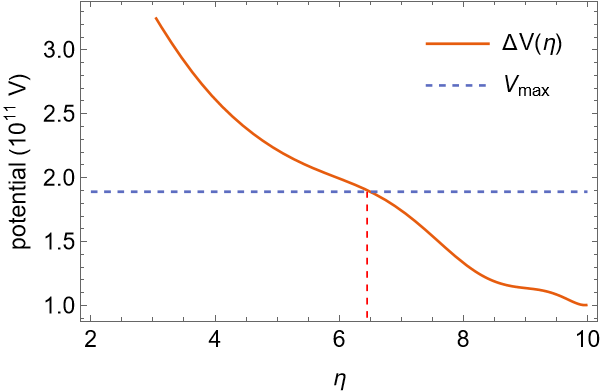}
    \caption{Orange solid line show the decreasing of the potential drop required by PSR J2144-3933 to form a spark as the  mountain steepness increasing, with blue dashed marking the maximum potential produced by unipolar induction. When $\eta>2.1$, it hold that $\Delta V<V_\text{max}$}
    \label{fig:star2}
\end{figure}

Applying the same methodology, we can explain the unexpected sparking behavior of PSR J2144-3933 \citep{Young+etal+1999} with a period of $8.51\:\mr{s}$ and a period derivative $4.96\times 10^{-16}\:\mr{s/s}$. According to the dipolar magnetic configuration, it has a surface magnetic field of $2.08\times 10^{12}\:\mr{G}$ and a energy loss rate of $3.2\times 10^{28}\:\mr{erg/s}$, with the spin down age of $272\:\mr{Myr}$. If there is a small mountain with a steepness $\eta$ greater than $6.4$ (Figure~\ref{fig:star2}), then it would be possible to generate sparks.
The parameters of these two pulsars and the changes in potential drops required to form sparks are concluded in Table~\ref{tab1}.

\begin{table}
\bc
\begin{minipage}[]{100mm}
\caption[]{Pulsar parameters and gap potential drop changes\label{tab1}}\end{minipage}
\setlength{\tabcolsep}{2.5pt}
\small
 \begin{tabular}{ccc}
  \hline 
  \noalign{\smallskip}
    Pulsar paramters 
    & J0250+5854 & J2144-3933 \\ \hline
    $P$ (s) & 23.54 & 8.51 \\
    $\dot{P}\; (\mr{s\,s^{-1}})$  & $2.72\times 10^{-14}$ & $4.96\times 10^{-16}$ \\
    $B$ (G) & $2.56\times 10^{13}$ & $2.08\times 10^{12}$ \\
     $T$ (K) & $9.86\times 10^5 $ & $4.2\times 10^5$ \\ \hline
    calculation results & & \\ \hline
    $\gamma_0$ & $1.76\times 10^5$ & $9.74\times 10^5$ \\
    $h_0$ (cm) & $3628.38$ & $9016.07$ \\
    $\Delta V_0$ (V) & $8.99\times 10^{11}$ & $1.25\times 10^{12}$ \\
     $V_\text{max}$ (V) & $3.04\times 10^{11}$ & $1.89\times 10^{11}$ \\
    $\eta_c$ & 2.1 & 6.4 \\
  \noalign{\smallskip}
  \hline
\end{tabular}
\ec
\centering
\tablecomments{0.86\textwidth}{The pulsar parameters used in this table is refered to \cite{Manchester+etal+2005}}

\end{table}

Despite the fact that the rotation period of PSR J0250+5854 is substantially greater than that of PSR J2144-3933, the magnetic field of the former is also significantly larger than that of the latter.
This results in the observation that, as illustrated in Figure~\ref{fig:ppdotm}, the former is closer to the pulsar death line and is able to generate sparks more easily by means of an amplified parallel electric field near mountains in the polar region. Therefore, it requires a smaller mountain steepness than the latter.

\section{Discussion}
\label{sect:discussion}
 
Apart from the revival of two ``dead" pulsars, the hypothesis of small mountains on the surface of the pulsar can explain other peculiar observational facts.
The observed offset between the main pulse and the inter-pulse emission peaks in PSR B0950+08 relative to the magnetic-axis line-of-sight plane implies surface magnetic field anomalies, suggesting a more intensive pair production activity above certain location of the polar cap surface, possibly indicative of multipolar components or crustal distortions \citep{Wang+etal+2024}. Moreover, mode switching process in some pulsars, e.g. PSR 0943+10 \citep{Cao+etal+2024}, may also reflect some changes on the stellar surface condition \citep{Bartel+1982, Vivekanand+1981}.
Furthermore, the diffuse drifting subpulses for PSR B2016+28 can be explained by a rough stellar surface (\citealt{lu+2019}), because the locally amplified accelerating field leads to a higher rate for pair cascade in the region surrounding the mountains, accounting for the abnormal and aperiodic deviation of the sparking discharge pattern.

The most important implication from those surface mountains is a constraint on the surface state of matter. Because of the extreme complexity of lattice quantum chromodynamics simulations non-perturbatively, it is currently impossible for us to theoretically determine the state of matter of pulsars. However, the presence of small mountains or other local uneven structures on the pulsar surface  requires the surface to have a strong shear modulus, or the thermal electrons would destroy the mountain in the gap, so the surface matter should be solid with strong shear modulus.

If a pulsar is a neutron star that is formed by the neutronization process as was proposed by Landau (\citealt{Landau:1932uwv}), the matter near the surface should be similar to normal matter combined by electromagnetic interaction due to continuity.
The binding energy is estimated to range from $10 \sim 100 \:\mr{eV}$, while thermal electrons in proximity to the surface possess kinetic energies ranging from $0.1 \sim 10\:\mr{keV}$. Thus, the mountains quickly collapse under the incessant bombardment of these high-energy electrons, resulting in a flat stellar surface. 

If a pulsar is a strangeon star proposed by \cite{Xu+2003} that takes into account the degrees of freedom of the quark, its surface would consist of strangeon matter, a condensate bound by the strong interaction with a binding energy of several MeV \citep{Xu+2023}, much higher than the thermal energy of surface electrons.  The relativistic ($\gamma\sim 10^{6}$) streaming particles that have been accelerated towards the stellar surface would collide with valence quarks in strangeons, but may not directly destroy the zit structure.
Consequently, the solid nature of the strangeon star's surface enables the stable existence of local unevenness.
Furthermore, from a symmetry-energy perspective, the strangeon phase is energetically favored over conventional neutron matter configurations \citep{Xu+2019}. Despite the maintenance of charge neutrality, the neutronization process is unable to preserve isospin symmetry concurrently.
However, the strangeonization process can simultaneously conserve both charge and isospin, thus exhibiting a higher degree of symmetry \citep{Xu+2021}.
Therefore, it may be more possible for the pulsar to take the form of a strangeon than a neutron star if it can be confirmed that mountains do exist on the stellar surface.

Notice some limitations of our current model. Our procedure to simulate the influence of small mountains in the polar cap region is built upon several hypotheses: (i) Magnetic axis and rotation axis are exactly anti-parallel for the purpose of simplifying the calculation.
However, it can still capture some critical features when the angle between two axes is small. (ii) The magnetic field configuration within the magnetosphere is dominated by dipolar magnetic fields, but the irregular structure on the surface is likely to induce multipolar fields and alter the distribution of the accelerating electric field.
(iii) We approximate the magnetosphere to be quasi-static, but according to \citetalias{Ruderman+etal+1975}, the magnetosphere does not necessarily co-rotate with the star at the same angular velocity, with a difference between them as
\begin{equation}
    \frac{\Omega^*}{\Omega(h)}\simeq 1+\frac{3 h^2}{R_s^2}
\end{equation}
where $\Omega^*$ is the stellar angular velocity and $\Omega(h)$ is the angular velocity of the magnetosphere as a function of gap height. However, our approximation is justified because only the angular velocity of stellar rotation ($\Omega^*$) is used in the calculation based on Equation~\eqref{eq:basic}, and this equation is guaranteed by a serious electrodynamic analysis presented in Appendix B of \citetalias{Ruderman+etal+1975}. (iv) When the Lorentz factor for positrons is up to $10^6 \sim 10^7$, the contribution of the thermal ICS process is not negligible.

There are also some possible approaches to test whether there are mountains in the polar cap region of pulsars.
Since it is necessary for the pulsars below the death line in Figure~\ref{fig:ppdotm} to have mountains on the surface to generate radio pulses, the observed distribution of the discharge points of those pulsars would have a higher probability of exhibiting irregular features than that of normal pulsars above the death line.
This observational difference provides a diagnostic tool: by searching for characteristic emissions from surface mountains, we may identify the mechanism by which topographical features on apparently dormant pulsars can restart particle acceleration through magnetospheric disturbances.

With the assumption of mountains on the stellar surface, it could be natural to understand the correlation between the peculiar emission features and pulsar timing anomalies in a wind-braking model~\citep{Tong+2013}, as has already been noted observationaly~\citep{Kramer+2006, Lyne+2010}.

The rotational energy of a pulsar could be lost through the combined effects of magnetic dipole radiation and particle winds~\citep{Xu+2001}, as verified by  numerical simulations~\citep{Contopoulos+2006}, and the wind braking mainly contributes to changes in timing behaviour at a higher order~\citep{Wang+2012,Tong+2013, Tong+2014}.
With a locally enhanced accelerating field, the flow of the particle wind may increase due to sparking around a mountain in the polar cap region, resulting in significant timing irregularity.

In addition, the observations of single-pulse fluctuation~\citep{2023NatAs...7.1235C,2025A&A...695A.203J,2025RAA....25k5004Y}, unusual arc-like structures~\citep{2007MNRAS.379..932M} and distinct core-weak patterns~\citep{2019MNRAS.484.2725B,2023MNRAS.520.4173W} could also result from sparking around small hills within the polar cap.
In any case, a detailed study of single pulses, especially with China's FAST, would undoubtedly be encouraged to reveal the natural roughness of the pulsar surface.

Last but not least, the effect of mountains may also be implicated from single pulses observations. As the ICS process proceeds, the high-energy electrons undergo gradual cooling, thereby suppressing the production of electron-positron pairs and resulting in a corresponding decline in radiation intensities. In contrast, the emergence of a single pulse occurs almost instantaneously. Consequently, the temporal evolution exhibits an asymmetric structure, typically manifesting as a bright primary pulse followed by a succession of gradually diminishing secondary pulses.

\section{Conclusions}

We propose a methodology that utilizes the \citetalias{Ruderman+etal+1975} model and the resonant ICS process to calculate the height of the inner vacuum gap, as well as the potential drop across the inner gap required for discharges, under scenarios involving either a flat surface or a surface with a small mountain in the polar cap region.
We apply this model to understand the radio emissions of the pulsars PSR J0250+5854 and PSR J2144-3933, both of which lie below the pulsar death line.
The presence of small mountains can reduce the voltage required to form a spark, thereby explaining the unexpected radio emissions observed from these two dead pulsars and the irregular distribution of discharge points observed in some other pulsars.
The existence of sustained surface mountains requires a solid-state stellar structure capable of maintaining such topological features against the bombardment of relativistic pairs and the gravity, suggesting that the pulsars might be made up of strangeon matter favored by symmetry.
%
The non-symmetrical sparking of PSR B0950+08~\citep{Wang+etal+2024} and the mode switches of PSR B0943+10~\citep{Cao+etal+2024} may hint at small mountains existing on pulsars' surface, but great efforts to find more observational evidence for mountain building are encouraged, particularly using China's FAST (Five-hundred-meter Aperture Spherical Telescope).

\normalem
\begin{acknowledgements}
This work is supported by the National SKA Program of China (2020SKA0120100) and the NSFC (No.12261141690 and No.12403058).

\end{acknowledgements}

\appendix
\section{Mean free paths}
\subsection{The mean free path of high-energy photons}
\label{apx:mfp_ph}

High-energy photons can produce electron-positron pairs in a strong magnetic field when the magnetic field component perpendicular to the photon¡¯s propagation direction is non-zero.
For photons with energy $h\nu > 2m_e c^2$,  the mean free path $l_p$ during propagation through such a field is given by \citep{Erber+etal+1966}

\begin{equation}
    l_p =\frac{2\pi \lambda_c}{\alpha}\epsilon_s \mr{K}_{1/3}^2\left(\frac{2}{3\chi}\right) , \quad \chi =\frac{\hbar \omega_s}{2m_ec^2}\frac{B_\perp}{B_q} =\frac{\epsilon_s\epsilon_B}{2}\frac{l_p}{\varrho} , \label{eq:lp}
\end{equation}

where $\lambda_c$ is the reduced Compton wavelength, $\alpha$ is the fine structure constant, $\mr{K}_{1/3}(x)$ is the modified Bessel function of the second kind, $\epsilon_s=\hbar\omega_s/(m_ec^2)$ is the energy of the photon in units of the rest energy of the electron, $B_\perp=(l_p/\varrho)B$ is the magnetic field component perpendicular to photon propagating direction. $\varrho $ is the radius of curvature at the location with the magnitude $\varrho \sim R_s = 10^6\:\mr{cm} $. $\epsilon_B=B/B_q$ is the dimensionless magnetic field with $B_q=m_e^2c^3/(e\hbar)=4.414\times 10^{13}\:\mr{G}$ the critical magnetic field strength. Solving out Equations~\eqref{eq:lp} simultaneously gives the photon mean free path as a function of photon energy and magnetic field strength $l_p(\epsilon_s, \epsilon_B)$.

If $\chi\ll1$, Equation~\eqref{eq:lp} can be approximated in a simpler form:
\begin{equation} 
    l_p = \frac{4.4 \lambda_c}{\alpha}\frac{B_q}{B_\perp} \exp( \frac{4}{3\chi}),
\end{equation}
as was mentioned in \citetalias{Ruderman+etal+1975} and \cite{Zhang+etal+1996,Zhang+etal+1997}. 
However, since $\chi$ can typically reach values up to 0.2, we adopt a more universally applicable approximation form given in Equation \eqref{eq:lp}.

\subsection{The mean free path of electrons and photons}
\label{apx:mfp_el}

As discussed in Section~\ref{apx:mfp_ph}, the resonant ICS process dominates high-energy photons production.
The scattered photons have dimensionless energy of $ \epsilon_s\simeq 2\gamma^2\epsilon (1-\beta \mu_i) $ after undergoing two fold Lorentz boosts, where $\gamma=1/\sqrt{1-\beta^2}$ is the Lorentz factor of the positrons involved in the ICS process.
$\mu_i=\cos \theta_i $ is the cosine of angle between the direction of incident photons and moving positrons, and $\epsilon$ is the dimensionless energy of incident thermal photons.

The mean free path of relativistic electrons or positrons in the ICS process can be approximated by \cite{Dermer+etal+1990}.
\begin{equation} 
    l_e \sim \qty[\int \sigma_\mr{eff} (1-\beta \mu_i)n_\text{ph}(\epsilon) \dd{\epsilon} ]^{-1} ,\label{eq:le} 
\end{equation}

where $n_\text{ph}(\epsilon)$ is the photon number distribution against energy, which is set as half the black body spectrum.  (only when the incident angle is smaller than the right angle can the photon be scattered into high-energy photons.)
\begin{equation} 
    n_\text{ph}(\epsilon) \dd{\epsilon} = \frac{4\pi}{\lambda^3_c}\frac{\epsilon^2}{\exp \qty(\epsilon/\epsilon_\mr{th})-1} \dd{\epsilon}  , \label{eq:nph} 
\end{equation}

where $\epsilon_\mr{th}=k_BT/(m_ec^2)$ is the dimensionless thermal energy and $\lambda_c$ the reduced Compton wavelength. The $\sigma_\text{eff}$ is the effective cross section of ICS
\begin{equation} 
    \sigma_\text{eff}=\frac{\sigma_T}{2}\qty[\frac{u^2}{(u+1)^2}+\frac{u^2}{(u-1)^2+a^2}] , \label{eq:sigma_eff}
\end{equation}

where $\sigma_T$ is the Thomson cross section and $u=\epsilon'/\epsilon_B$ is the ratio of photon energy in positron rest frame $\epsilon'$ to cyclotron energy $\epsilon_B$, $a=2\alpha \epsilon_B/3$ \citep{Xia+etal+1985,Daugherty+1989,Dermer+etal+1990} is the resonance width.

Substituting Equation~\eqref{eq:nph} and Equation~\eqref{eq:sigma_eff} into Equation~\eqref{eq:le}, we get the positron mean free path as a function of energy of incident photons and  relativistic positrons and magnetic field strength 
\begin{equation} 
    l_e(\epsilon_s,\epsilon_B,\gamma) = \qty[\frac{\sigma_\text{eff}(\epsilon_s,\epsilon_B)}{2\pi^2\lambda_c^3}\frac{\epsilon_B}{\gamma} f(\epsilon_s,\gamma) ]^{-1}, \label{eq:lefunc}
\end{equation}

where $f(\epsilon_s,\gamma)$ is an integration related to energy of positrons and photons involved and takes the form below
\begin{equation} 
f(\epsilon_s,\gamma) = \int_{\epsilon_0}^{\epsilon_+}\frac{\epsilon}{\mr{e}^{\epsilon/\theta}-1} \dd{\epsilon} .
\end{equation}

It can be analytically worked out with poly logarithm functions, with $\epsilon_0=\epsilon_s/(2\gamma^2)$ corresponding to energy of incident photons that run perpendicular to positrons and $\epsilon_+=\epsilon_s/(2\gamma^2(1-\beta))$ the photons run just the same direction of positrons.

Figure~\ref{fig:free-path} shows the relation between the mean free path of positrons and photons to the energy of up-scattered photons in the polar gap region for a typical pulsar with $B=10^{12}\:\mr{G},P=1\:\mr{s}, T=10^6\:\mr{K}$ and the Lorentz factor of positrons $\gamma=10^5$, from which we can observe a sharp dip of the mean free path of the positrons at the resonance energy of magnetic field $\epsilon_s=2\gamma \epsilon_B=2\gamma \hbar eB/(m_ec)$.
This sharp dip originates from the Breit-Wigner-like distribution of the effective cross section Equation~\eqref{eq:sigma_eff}. The resonance condition reads that $u=1$ in Equation~\eqref{eq:sigma_eff}, which can thus be approximated by 
\begin{equation} 
    \sigma_\text{eff}=\frac{\sigma_T}{2}\qty[\frac{1}{4}+\frac{1}{\delta^2+a^2}], 
\end{equation}

with $\delta=a\tfrac{1+a^2}{1-a^2}$ the half width at half minima of Breit-Wigner distribution.

\bibliographystyle{raa}
\bibliography{bibtex}

\end{document}